\documentclass[acmlarge]{acmart}

\AtBeginDocument{%
  \providecommand\BibTeX{{%
    \normalfont B\kern-0.5em{\scshape i\kern-0.25em b}\kern-0.8em\TeX}}}

\setcopyright{acmcopyright}
\copyrightyear{2020}
\acmYear{2020}

\acmJournal{DTRAP}
\acmVolume{00}
\acmNumber{0}
\acmArticle{000}
\acmMonth{12}

\usepackage{multirow}

\begin{document}

\title{Treadmill Assisted Gait Spoofing (TAGS): An Emerging Threat to Wearable Sensor-based Gait Authentication}

\author{Rajesh Kumar}
\orcid{0000-0001-7467-5762}
\affiliation{%
  \institution{Syracuse University and Haverford College}
  \country{USA}}
\email{rkuma102@syr.edu}

\author{Can Isik}
\affiliation{%
  \institution{Syracuse University}
  \country{USA}}
\email{cisik@syr.edu}

\author{Vir V. Phoha}
\affiliation{%
  \institution{Syracuse University}
  \country{USA}}
\email{vvphoha@syr.edu}

\renewcommand{\shortauthors}{Kumar et al.}

\begin{abstract}
In this work, we examine the impact of Treadmill Assisted Gait Spoofing (TAGS) on Wearable Sensor-based Gait Authentication (WSGait). We consider more realistic implementation and deployment scenarios than the previous study, which focused only on the accelerometer sensor and a fixed set of features. Specifically, we consider the situations in which the implementation of WSGait could be using one or more sensors embedded into modern smartphones. Besides, it could be using different sets of features or different classification algorithms, or both. Despite the use of a variety of sensors, feature sets (ranked by mutual information), and six different classification algorithms, TAGS was able to increase the average False Accept Rate (FAR) from 4\% to 26\%. Such a considerable increase in the average FAR, especially under the stringent implementation and deployment scenarios considered in this study, calls for a further investigation into the design of evaluations of WSGait before its deployment for public use.
\end{abstract}



\maketitle

\section{Background and Motivation} \quad 
Over a hundred studies have been published over the past two decades on Wearable Sensor-based Gait {Authentication} (WSGait) \cite{sprager2015inertial, PhohaSurvey, RossSurvey2018, NIteshOffenseAndDefenseSurvey, VishalSurvey, GaitWearableSurvey2019}. Most of these studies have evaluated the uniqueness of WSGait under the zero-effort attack scenario. Only a handful of studies have paid attention to circumvention of WSGait via imitation (does not include data injection) \cite{Gafurov2006, GafurovGender2007, GafurovGender2009DifferentPositions, Gafurov2007TIFS, StangAttack2007, Mjaaland2011, TreadmillAttack,LatestimitationFail}. To the best of our knowledge, all but two \cite{StangAttack2007, TreadmillAttack} of these studies have shown that it is non-trivial to circumvent WSGait via imitation. With a limited experimental setup, Kumar et al. \cite{TreadmillAttack} showed that one could spoof WSGait with comparative ease by adjusting gait characteristics such as speed, step-length, step-width, and thigh-lift with the help of an off-the-shelf treadmill. Because treadmills are easily accessible to anybody, the treadmill-assisted attack exposes a vulnerability, which hitherto {was} thought to be non-trivial poses an emerging threat to WSGait as we expect WSGait to become one of the possible means of authentication in the future. The scope of \cite{TreadmillAttack} is limited to the accelerometer sensor and a fixed set of features. Furthermore, the same set of features was used to train the imitator, which could have favored the attack. Therefore, the motivation {is} to explore the applicability of treadmill attack under more stringent scenarios, including those in which WSGait would be using readings of one or more sensors embedded into the device and a different feature set than the one used for training the imitator.

\section{Related Work}
\quad This work primarily extends the idea presented in \cite{TreadmillAttack}. The notable differences between this and the previous work \cite{TreadmillAttack} are as follows:

\begin{itemize}
    \item  The authentication systems tested for circumvention in \cite{TreadmillAttack} were built using only one sensor{'s} readings, i.e., accelerometer readings. In practice, however, the authentication system could be using readings of multiple sensors embedded in smart-devices \cite{Lee2015MultisensorAT, PhoneMovementPrinceton}. Previous studies have reported that the fusion of multiple sensors helps achieve better error rates \cite{Lee2015MultisensorAT, PhoneMovementPrinceton, Smartwatch2015}. Thus, it becomes a logical next step to test the effectiveness of the attack presented in \cite{TreadmillAttack} on systems that use the readings of more than one sensor for user authentication. Since it is not always possible to find out the sensors in use, we decided to study four commonly used sensors (accelerometer, gyroscope, magnetometer, and rotation vectors) and their all possible combinations \cite{Lee2015MultisensorAT, PhoneMovementPrinceton, Smartwatch2015, SmartWatchWisdom}.
    \item The attack presented in \cite{TreadmillAttack} uses the same set of features for both, (1) training the imitator and (2) training and evaluating the authentication models. In other words, although unstated, the authors of \cite{TreadmillAttack} assume that the attacker would know the set of features used to implement the authentication system subject to the attack. In practice, this assumption may not be valid. The knowledge of the feature space makes the model even more susceptible to attack. For example, Zhao et al. \cite{FeatureLengthRandomAttackNDSS2020} showed that the knowledge of the length of the feature space helps launch a data injection attack on WSGait. Therefore, it is likely that each user authentication model is implemented using a different set of features selected by a feature selection algorithm (see Figure \ref{FlowchartWSGait3}). Therefore, a logical extension of the attack presented in \cite{TreadmillAttack} is to test its effectiveness on the authentication systems that may or may not be using the same set of features that the attacker used to train the imitator.
    \item {We elaborate on the working of the attack}, in what scenarios it applies, how to access the required information needed to launch the attack and improve its success, advantages over other attack methods, limitations, and possible countermeasures.

\end{itemize}
\quad Besides \cite{TreadmillAttack}, the closely related works are listed in the Table \ref{tab:RelatedWorks}. We can see that the implementation strategies of mimicry attacks have evolved over the past decade. Gafurov et al. \cite{Gafurov2006, GafurovGender2007, GafurovGender2009DifferentPositions, Gafurov2007TIFS} investigated the security of gait under two types of deliberate spoofing scenarios, friendly- and hostile. Under the friendly-scenario, the imitators did not make any deliberate attempt to copy the genuine user. While under the hostile-scenario, individuals with similar physical characteristics attempted to imitate each other. The chances of impostor acceptance did not increase in either of the scenarios. The authors concluded that the imitator with similar physical characteristics or of the same gender might have a better chance of getting accepted \cite{GafurovGender2007}.

In a follow-up, Stang et al. \cite{StangAttack2007} analyzed five templates of the same user collected at different settings of gait-factors such as speed, step-length, etc. The authors employed a total of 13 imitators; each made fifteen attempts to match each template. The whole experiment took about thirty minutes. The imitators had not seen the target walking before they started the imitation process. Live plots of resultant accelerations were shown to the imitators on a big screen in addition to the match scores (Pearson's correlation coefficient). The match score was computed between the patterns of the imitator and the target user. Some imitators were able to exceed the correlation coefficients of 0.5 (considered a 50\% match) criteria set by the authors, but randomly which led to the conclusion that one may imitate the gait patterns of targeted individuals if trained rigorously. Later, Mjaaland et al. \cite{Mjaaland2011} pointed out that the conclusion of Stang et al. \cite{StangAttack2007} cannot be relied upon at least for two reasons. First, the authors studied only five gait templates that {were} collected from only one individual. Second, Pearson's correlation coefficient between the resultant acceleration of the target and the imitator samples was used to measure the attack's success.

\begin{table}[htp]
\caption{List of most closely related work, publication year, number of subjects, gait cycles per subject, imitators per subject along with the number of cycles, error under the zero- and mimicry-effort attacks. The size of the data is provided in terms of number of gait-cycles to make a comparison among different works easy. In this work, however, we used a sliding window-based segmentation approach which output fixed length frames (see Section \ref{PreprocessingAndSegmentation}). \textit{Each of the frames was ten seconds long and roughly constitutes about seven gait-cycles considering that an average individual takes about 1.4 seconds to complete a gait cycle \cite{GaitFactorValues,GaitFactorValues2}}.}
\small
\centering
\label{tab:RelatedWorks}
\begin{tabular}{|l|l|l|l|l|l|l|}
\hline
Authors         & Year    & \begin{tabular}[c]{@{}l@{}}Users \\ (cycles/user)\end{tabular} & \begin{tabular}[c]{@{}l@{}}Imitators per \\ user (cycles)\end{tabular} & \begin{tabular}[c]{@{}l@{}} Avg. \\ Error (\%) \\ (before)\end{tabular} & \begin{tabular}[c]{@{}l@{}} Avg. \\ Error (\%) \\ (after)\end{tabular} & Remarks                                                              \\ \hline
Gafurov et al. \cite{Gafurov2006} & 2006    & 22 (15)                                                    & 1 (2)                                                                  & 16                                                             & 16                                                            & No substantial threat                                                          \\ \hline
Gafurov et al. \cite{GafurovGender2007} & 2007    & 100 (4)                                                   & 1 (4)                                                                  & 13                                                             & 13                                                            & Same-gender a threat                                                   \\ \hline
Gafurov et al. \cite{Gafurov2007TIFS} & 2007    & 100 (4)                                                   & 1(4)                                                                   & 13                                                             & 13                                                            & \begin{tabular}[c]{@{}l@{}} Same-gender \\ or closest in dataset is a threat\end{tabular} \\ \hline
Stang et al. \cite{StangAttack2007}    & 2007    & 1 (5)                                                     & 1(13)                                                                  & 26                                                             & --                                                            & \begin{tabular}[c]{@{}l@{}}Correlation for assessing success, \\ only one target user \end{tabular}                                                 \\ \hline
Mjaaland et al. \cite{Mjaaland2011} & 2011 & 50 (10)                                                   & --                                                                       & 6.2                                                            & 6.2                                                           & \begin{tabular}[c]{@{}l@{}}Training makes it worse \end{tabular}                                 \\ \hline
Kumar et al. \cite{TreadmillAttack}   & 2015    & 18 (532)                                                  & 1(970)                                                                 & FAR:6                                                          & SFAR:38                                                         & \begin{tabular}[c]{@{}l@{}}Controlled training on a treadmill\end{tabular}                                                          \\ \hline
Muaaz et al. \cite{LatestimitationFail}   & 2017    & 35 (200)                                                  &              \begin{tabular}[c]{@{}l@{}}5 victims \& \\ 4 imitators\end{tabular}                                                          & 13                                                             & 13                                                            & \begin{tabular}[c]{@{}l@{}}Training makes it worse, \\ used actors\end{tabular} \\ \hline
This work       & 2020    & 18 (532)                                                  & 1(970)                                                                &  FAR:4                                                              &                      SFAR:26                                         & \begin{tabular}[c]{@{}l@{}}Multiple sensors, Frame-based, and \\ Dynamically selected features for \\ each authentication model \end{tabular} \\ \hline
\end{tabular}
\end{table}

Mjaaland et al. \cite{Mjaaland2011, MjaalandAttack2010Plateau, Mjaaland2009GaitMAThesis} also pointed out that the previous studies merely moved beyond minimal-effort mimicry, which led to the investigation of what impact an extensive training would have on the imitation process. The authors hired a total of seven imitators. Six of which were trained for more than an hour. The seventh imitator was trained for as long as six weeks. The systematic change in the learning curve of imitators was observed using linear regression. Specifically, $Y (i) = \beta_1 + \beta_2 \times e^{\beta_3/i}$ for observation $i$ where $\beta_1, \beta_2,$ and $\beta_3$ are constants, a non-linear (in $\beta_3$) regression model was used \cite{ProbabilityStats}. The residual error for the model was defined as $r_i = |Y(i)-y(i)|$, where $y(i)$ is the value at $i$th observation. The constants $\beta_2$ and $\beta_3$ and their magnitude indicated the progress and rate of the (un)learning, respectively. The same signs (a downward sloping curve) of these constants implied that the imitators' learning improved. On the other hand, opposite signs indicated that the imitators' learning worsened. 
The authors conducted the experiments under three scenarios, friendly, short-term hostile, and long-term hostile scenarios. \textit{The friendly scenario} consisted of collecting regular walking patterns from fifty participants while they walked a fixed distance without any training or feedback. Each participant was filmed while walking. This scenario was considered a baseline scenario in the study and resulted in an Equal Error Rate (EER) of 6.2\%. \textit{In the short term hostile scenario}, one target and six imitators were selected from the fifty participants. The criteria for choosing the six imitators included was the stability of their gait (low intra-variation), the Dynamic Time Warping (DTW) distance of their gait patterns from the target's gait patterns, and their eagerness to participate in the study. Imitators of both types, whose gait patterns were too close and whose gait patterns were too far from the target, were selected.

Each of the imitators made five attempts, each lasting about an hour. In each session, the video clips of the imitation attempts and distance scores were shown to the imitators. Possible improvements were discussed/suggested before making the next attempt. In the first two sessions, high-level information such as speed, sideways postures, and arm swing was tuned. In the third session, the imitators walked behind the target. This attempt was also recorded and was one of the most valuable (reported by imitators) feedback. Specifically, this attempt helped to synchronize the speed and step length. Session four was focused on finer details such as feet, ankle, heel, toe, hip, torso, and shoulder movements. Session five was the final imitation trial without any training or feedback. Finally, in \textit{the long-term hostile scenario}, one imitator who was available for at least six weeks was chosen. The imitator made a total of sixty attempts, which took about sixty hours, ten hours per week.

The authors concluded that all six imitators could not breach their respective physiological boundaries. The boundaries were visible for each imitator on the graph plotted from the regression analysis. The authors observed that the imitators could adapt specific characteristics of the target but ultimately failed to match all of the traits. The learning curve for some imitators worsened over time. The imitators reported that over-training was making their walk more unnatural and mechanical. The long-term scenario indicated that a person could have multiple plateaus but with a great deal of uncertainty. Various plateaus produced by the same imitator suggested that an imitator can match the target. Still, due to a high level of uncertainty and insufficient data, the authors could not conclude anything.

Interestingly, the authors pointed out that statistical feedback was much more helpful to the imitators than visual feedback. For some attackers, the visual appearance of the walk was almost similar to that of the target. But the data collected from those attempts were far away (DTW) from that of the target patterns. The participating imitators suggested that the most challenging part was to concentrate on the target's different gait factors simultaneously. Besides, they kept forgetting what they had learned in the previous attempt. Ultimately, the authors concluded that gait imitation is challenging to achieve.

Later, Muaaz and Mayrhofer \cite{LatestimitationFail} studied WSGait spoofing in two phases, \textit{reenact} and \textit{coincide}. The imitators observed the target user for about ten minutes while walking next to them in the \textit{reenact} phase. In the \textit{coincide} phase, real-time feedback was computed by comparing the gait patterns of the target and the imitator. Nine individuals, specifically trained in mimicking body motions and body language, participated in the process. Five of the nine individuals acted as imitators and the rest as victims. None of the imitators could produce gait samples that would match with those of the targeted individuals. The authors finally concluded that circumventing WSGait via spoofing is a difficult task \cite{LatestimitationFail}.

The experimental setups of the above studies imply two underlying hypotheses. First, two visually similar walks from different individuals would produce matching sensor readings or feature values extracted from the sensor readings. Second, human imitators would learn, adapt, and repeat the targeted gait patterns at will. In contrast, the authors of \cite{TreadmillAttack} focused on matching features extracted from the sensor readings produced by the target. They further stated that human imitators might be able to learn, adapt, and repeat the targeted gait patterns more easily with the assistance of a device like a treadmill. The attack based on these ideas achieved an average spoof false accept rate (SFAR) of $43.66\%$ for accelerometer-based gait {authentication} over a dataset of eighteen users and one imitator \cite{TreadmillAttack}. Hence the motivation to explore the idea of \cite{TreadmillAttack} further under improved scenarios. 

Besides the works mentioned above, some recent studies attempted circumvention of WSGait by intercepting the authentication pipeline and injecting synthetic data \cite{OneCycleAttack, VideoGANReviewer3}. For example, Jia et al. \cite{VideoGANReviewer3} utilized Silhouette Guided Generative Adversarial Networks (GANs) to spoof video-based gait recognition systems. Zhu et al. \cite{OneCycleAttack} used k-means clustering to generate a variety of attack patterns and injected them into the pipeline to successfully fool the sensor-based gait-authentication system. Our study differs from these studies as we focus on circumvention via imitation while they investigate circumvention via data injection \cite{OneCycleAttack,FeatureLengthRandomAttackNDSS2020}. Please refer to Section \ref{AdvantageOverDatainjection} for more details.

\section{Design of Experiments}\label{DesignOfExperiments} 
\subsection{Typical implementation frameworks for WSGait}
\quad A typical implementation of WSGait includes data acquisition via sensors, preprocessing and segmentation, feature extraction and selection, and training and authentication (see Figure \ref{FlowchartWSGait3})\cite{sprager2015inertial}.

\subsubsection{Data acquisition via sensors} 
A variety of wearable sensors, including accelerometer, gyroscope, and magnetometer available into smart-devices, can be used to capture individuals' gait patterns. The accelerometer is considered one of the best sensors to capture the uniqueness of the individual's gait patterns \cite{Lee2015MultisensorAT}. Besides, the combination of multiple sensors offers better authentication accuracy \cite{PhoneMovementPrinceton}.
\subsubsection{Preprocessing and segmentation}
\label{PreprocessingAndSegmentation} Mechanisms such as moving average and multi-level wavelet decomposition and reconstruction are used to remove noise and outliers from sensor readings \cite{MjaalandAttack2010Plateau}. Considering \textit{low computational complexity} and \textit{low recognition latency} as some of the important design goals, the sensor readings are divided into small segments, either gait-cycles or frames with a fixed length. Commonly used approaches for detecting gait-cycles include local extrema analysis \cite{Gafurov2006, GafurovGender2007, Gafurov2007TIFS}, zero-crossing detection, and phase analysis, among others. The detected cycles are further aligned and normalized for point-wise comparison. On the other hand, frames with fixed length are obtained either in an overlapping or non-overlapping manner. Overlapping frames can be obtained by using a sliding window mechanism \cite{Smartwatch2015}. As per previous studies, \cite{Smartwatch2015, PhoneMovementPrinceton} frame length between 8-12 seconds and overlap of half of the frame length offer better recognition. A list of features is computed from each frame, and the resultant set of frames is fed to machine learning algorithms for training authentication model. Many state-of-the-art implementation approaches of WSGait rely on frame-based segmentation coupled with machine learning algorithms \cite{sprager2015inertial,PhoneGait1,damavsevivcius2016smartphone,PhoneMovementPrinceton, Lee2015MultisensorAT}.

\subsubsection{Feature extraction and selection} The existence of this component depends upon the type of matching strategy one follows. For example, no feature extraction or selection is required in case the matching involves a direct comparison of extracted segments using distance measures such as DTW \cite{muaaz2013analysis, PhoneGait2,muaaz2012analysisWeakDTW, LatestimitationFail} and distance metrics such as Euclidean \cite{Gafurov2006, GafurovGender2007, Gafurov2007TIFS}. The direct comparison usually applies to cycle-based segmentation and is referred to as a template-based approach. However, several recent studies use a fixed-length frame approach. The authors extracted a series of features from the raw sensor readings and employ a variety of machine learning algorithms to train authentication models \cite{ PhoneGait7,damavsevivcius2016smartphone, SVMImpBorrowedFrame, PhoneMovementPrinceton, Lee2015MultisensorAT,kNNForGait}. In some cases, e.g., \cite{TreadmillAttack}, authors select the most distinguishing features using mutual information between each feature and the class label.

\subsubsection{Training and authentication} The training component primarily involves the training of the authentication model using classification algorithms or the creation of a template for each user from the data collected during the training session. The \textit{authentication} component uses the trained authentication model or the template to classify the incoming test segments (or feature vectors) into genuine or an impostor. For training two-class classification algorithms, one needs samples belonging to both genuine and impostor classes. Authors often use samples collected from other users than the genuine user as impostors \cite{SVMImpBorrowedFrame, PhoneMovementPrinceton, Lee2015MultisensorAT, TreadmillAttack}.

\subsection{Choice of the implementation framework}
\begin{figure}[htp]
\centering
\includegraphics[width=5.6in, height=1.7in]{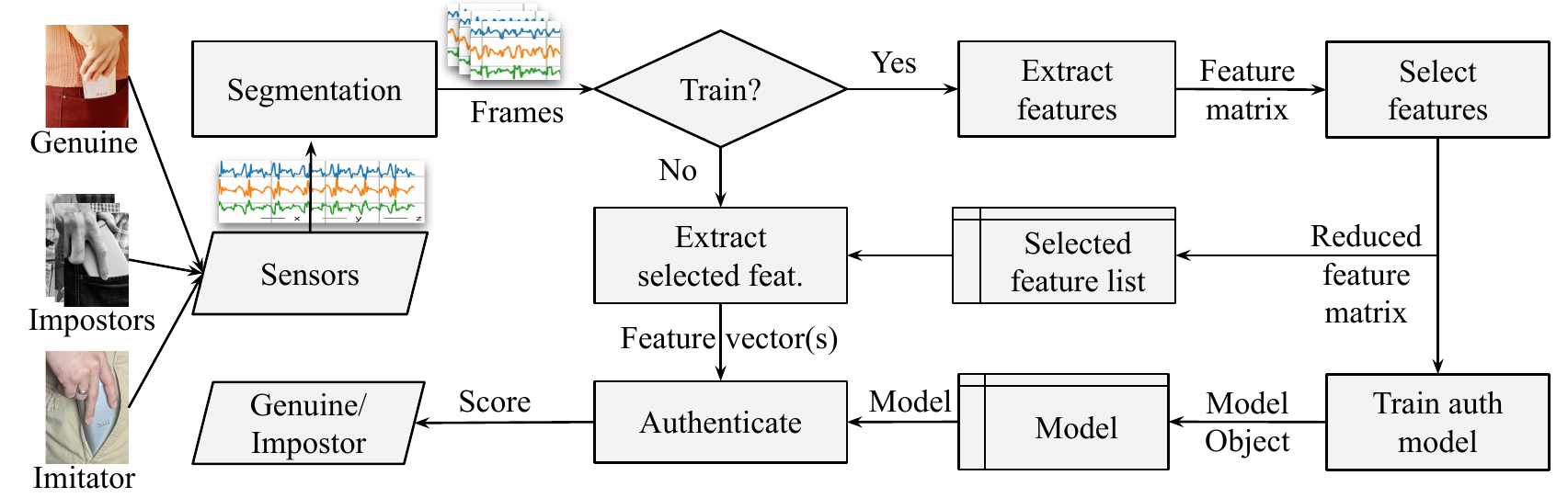}
\caption{The pipeline used for training the authentication system of each user. \textit{Genuine} represents those users for whom the authentication system is being trained. \textit{Impostors} {applies} to every user in the database except the Genuine user. It is important to note that, the impostors do not make any deliberate attempt to mimic the \textit{Genuine} user; their regular gait patterns were used as impostor samples following the past studies \cite{SVMImpBorrowedFrame, OneClassISABA2018, TreadmillAttack}. On the other hand, \textit{Imitator} is the user who made deliberate attempt to copy the \textit{Genuine} user after receiving feedback-based training (see Section \ref{AttackImplementation} for more details).}
\label{FlowchartWSGait3}
\end{figure}

The fixed-length frame-based approaches \cite{PhoneGait7,damavsevivcius2016smartphone,SVMImpBorrowedFrame,PhoneMovementPrinceton,Lee2015MultisensorAT,kNNForGait} coupled with machine learning algorithms (e.g., k-nearest neighbors, support vector machines, neural network etc.) usually outperform cycle-based approaches that use direct comparison of segmented data points using distances metrics (e.g., Euclidean or Manhattan) or distance measures such as DTW \cite{muaaz2013analysis, PhoneGait2,muaaz2012analysisWeakDTW,LatestimitationFail,Gafurov2006, GafurovGender2007, Gafurov2007TIFS} in the context of authentication. Thang et al. \cite{PhoneGait7} specifically investigated both approaches and concluded that the former approach achieved significantly better results $(92.7\%)$ than the latter $(79.1\%)$ on the same dataset. Besides, a recent study by Al-Naffakh et al. \cite{FrameIsBetterThanCycleInGait} presented a qualitative comparison of 29 studies and concluded that the former approach almost always beats the latter. For this work, we implemented both approaches. We found that the latter approach exhibited very high ($\ge 20\%$) error rates on the test data (collected two to three days later) compared to the former approach that consistently achieved average error rates below $7\%$ (see Zero-effort HTER heatmap in Figure \ref{AttackImpact}). The high error rates exhibited by the latter approach could be an outcome of several factors. The factors include low sampling frequency in the dataset, relatively more realistic experimental environment (the participant kept the phone in their pant pocket), and the substantially higher number of gait cycles per user (see Table \ref{tab:RelatedWorks}). We posit that the authentication system that exhibits more than 20\% error rates on the test data would not be suitable for most of the scenarios (e.g., securing expensive cars, houses, offices) in which WSGait is applicable. Needless to mention that the models implemented using the latter approach may suffer from even higher error rates if deployed for use in their current form. Therefore, we decided to focus on WSGait implemented using the former approach. Figure \ref{FlowchartWSGait3} depicts the overall experimental setup.

\subsubsection{Dataset} The dataset used in this study was originally collected and used by Kumar et al. \cite{TreadmillAttack}. Although the dataset consisted of readings from four sensors, Kumar et al. \cite{TreadmillAttack} utilized only accelerometer sensor readings, whereas we use the readings of all four sensors. The essential details of the data collection experiment are provided below.

The dataset consists of gait patterns collected from eighteen \textit{Genuine} users and one \textit{Imitator} who mimicked each of the Genuine users one by one. The \textit{Genuine} users walked naturally in a $100 \times 2$ $meter^2$  corridor back and forth for about two minutes keeping an HTC-One M8 smartphone in the right pocket of their pants. The phone was always positioned upside-down, and the screen faced the participants' bodies. The particular setup was adopted to keep the data collection environment uniform and consistent among the users. This data collection setup, however, overlooks the scenarios in which the users can place their phone in any position (e.g., upside-up) and location (left or back-pocket) on the body. For more details, please refer to Section \ref{Limitations}. 

The readings of four sensors, namely, accelerometer, gyroscope, magnetometer, and rotation vector, were captured by an App installed on the phone. The \textit{Genuine} users repeated the exercise after two to three days. The data collected in the first phases were used to train and validate the model; thus, it would be referred to as the \textit{Training dataset}. In contrast, the data collected in the repeat session was used for evaluating the model; hence, it would be referred to as \textit{Testing dataset} in the rest of this document. On the other hand, the data collected from the trained \textit{Imitator} (see Section \ref{AttackImplementation}) shall be referred to as \textit{Mimicry dataset} in the rest of this document.

\subsubsection{Preprocessing, segmentation, feature extraction, and feature selection}
\label{FeatureAanlysis}Following previous studies \cite{Mjaaland2009GaitMAThesis,Mjaaland2011,MjaalandAttack2010Plateau}, {a} simple moving average technique was applied to filter the noise. Then, the sensor readings were segmented using a sliding window mechanism with a frame length of ten seconds and an overlap of five seconds. The segmentation process resulted in a set of independent frames. On average there were twenty frames per user in the \textit{Training} and \textit{Testing} dataset. While there were sixty-six frames (collected in three independent attempts) for each user in the \textit{Mimicry} dataset. Adapting from previous studies \cite{PhoneGait1, PhohaSurvey, PhoneMovementPrinceton, Abena}, we extracted a series of time and frequency domain features from each of the frames. The specific time-domain features include arithmetic mean; standard deviation; mean absolute change; mean absolute deviation; skewness; kurtosis; mean energy; the number of mean crossings; the number of peaks; first, second, and third quantiles; length of the longest strike below and above the mean; and bin counts in $16$ equally thick bins. The bin counts were included primarily because of its effectiveness reported by some of the seminal studies \cite{PhoneGait1, SmartWatchWisdom}, and it provides different information compared to aggregate features. The frequency-domain features included the first, second, third quantiles, and standard deviation of the Fourier transform's amplitudes.

Each sensor used for data collection has three components $x$, $y$, and $z$, i.e., three time-series signals. The fourth signal, i.e., magnitude, was computed as $m =\sqrt{x^2+y^2+z^2}$ and considered the fourth component. The features were extracted from each component and concatenated together. As a result, we had 136 features from each sensor. Those were too many features to work with, especially if we want to develop a lightweight system.
Consequently, following previous studies \cite{HMOG, TypingSwipingFusion,PhoneSwiping1}, we included mutual information (MI) based feature selector in the pipeline to select the top $k$ most discriminating features. We did not want to have too few or too many features; we experimented with seven different values (start=20, end=50, step=5) of $k$ and decided to use $30$ in the end, as it provided the best results under the Zero-effort testing environment. The feature selector computed MI between each feature and the class label and output the top $k$ features with the highest MIs. We observed that the chosen features varied for different users but not significantly.

\subsubsection{Choice of classification algorithms} The next step was to choose classifiers to train the authentication models. Following previous studies \cite{OneClassISABA2018, PhoneGait7, Abena,Nickel2011SVMFrames,ChungXu,PhoneMovementPrinceton,PhoneGait1,SVMImpBorrowedFrame}, especially \cite{TreadmillAttack}, we employed Bayes Network (Bayes), Logistic Regression (LogReg), Multilayer Perceptron (MulPer), Random Forest (RanFor), and Support Vector Machines (SVM) to classify the feature vectors between genuine and impostors. Besides these classifiers, we included k-Nearest Neighbors (kNN) because previous studies \cite{ChungXu, kNNForGait, OneClassISABA2018} have demonstrated its superiority over the other approaches. To be trained, these classifiers required feature vectors from both genuine and impostor classes. The Genuine feature vectors were created from genuine user's data, while the impostor feature vectors were created from data belonging to Impostors (see Figure \ref{FlowchartWSGait3}). Since there were seventeen impostors, we used about 30\% (six) feature vectors from each, which resulted in a total of $108$ impostor feature vectors. Since we had only 20 genuine feature vectors for each genuine user on average, this situation turned into a class imbalance problem. The problem was addressed by oversampling the Genuine feature vectors using Synthetic Minority Oversampling (SMOTE) \cite{SMOTE}, a widely used method in the field. The SMOTE application helped balance the number of feature vectors in each class and increased the number of feature vectors used for training to $216$. \textit{As a result, the training matrix for each user model was of size $216 \times 30$ for each user authentication model, where $30$ is the number of features.}

\subsubsection{Hyper-parameter tuning and performance evaluation} The hyper-parameter was tuned by conducting 10-fold cross-validations on the \textit{Training} dataset using Half Total Error Rate (HTER) as a loss function \cite{BengioHTERLossFun}. HTER is defined as an average of False Accept Rate (FAR) and False Reject Rate (FRR). The authentication models were evaluated on the feature vectors extracted from the \textit{Testing} and \textit{Mimicry} datasets in two phases. The first phase focused on Genuine testing and was conducted on the genuine feature vectors. The second phase consisted of two parts, \textit{impostor testing} and \textit{imitator testing}. The \textit{impostor testing} was conducted on the impostor feature vectors, which were created from the users other than the genuine users in the \textit{Testing} dataset. This testing environment is referred to as the \textit{Zero-effort} attack environment. On the other hand, the \textit{imitator testing} was conducted on the feature vectors extracted from the data available in \textit{Mimicry dataset} for each genuine user. This testing environment would be referred to as the \textit{Treadmill-assisted} attack environment.

Each of the feature vectors was assigned either genuine (accept) or impostor (reject) class labels during testing. One can devise a policy on how many consecutive accepts or rejects granting or denying access can be devised for different application scenarios. For example, one can secure a smart-car via WSGait. The smart-car doors may open only after the user produces five continuous matching frames. For simplicity, like previous researchers, we report the performance of WSGait in terms of False Accept Rate (FAR) and False Reject Rate (FRR), which are defined as follows:

\quad \textit{FAR = number of impostor frames that were accepted / number of total impostor frames}

\quad \textit{FRR = number of genuine frames that were rejected / number of total genuine frames}\\
FAR and FRR represent Impostor pass and Genuine fail rates, respectively. Similar to FAR (the impostor pass rate), we will use the Spoof False Accept Rate (SFAR) to evaluate the Imitator pass rate. Also, we provide Half Total Error Rate (HTER) and Spoof Half Total Error Rate (SHTER). The HTER is nothing but an average of FAR and FRR. Likewise, the SHTER is an average of SFAR and FRR \cite{HMOG}.

\begin{figure}[htp]
\centering
\includegraphics[width=5.6in, height=2.15in]{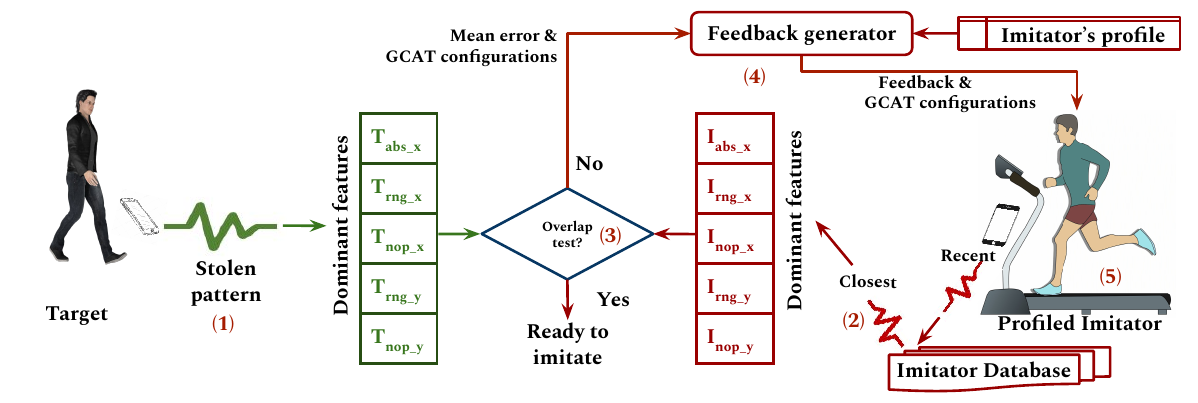}
\caption{The human imitator (on the right) is trying to produce accelerometer readings that would closely match with that of the target user (on the left) at the feature level. (n) denotes the step \# n. The feedback loop (2-3-4-5-2) ends once each of the five features passes the overlap test.}
\label{FeedbackAttackProcess}
\end{figure}

\subsection{The imitation process} \label{AttackImplementation} \quad The imitation process is illustrated in Figure \ref{FeedbackAttackProcess}. The process began by creating an imitator's profile and was followed by training the imitator to mimic the target's gait patterns. The imitator profile was a two-dimensional correlation matrix. The coefficients were computed between the dominant feature set, which consisted of five elements viz. abs\_x, rng\_x, nop\_x, rng\_y, nop\_y and the Gait Characteristics Adaptable over a Treadmill (GCAT). The abbreviations abs, rng, and nop refer to absolute sum, range, and the number of peaks, respectively. 

The GCAT included speed, step-length, step-width, and thigh-lift. The data used for computing the correlation was collected at different configurations of GCAT. The speed's configuration values were between 1.2 and 2.8, with an increment of 0.2 miles per hour. The configuration values for the step-length included small, regular, and large. Similarly, the step-width configurations had close, regular, and wide, and the thigh-lift included front, regular, and back. While changing the configuration values for one characteristic, the levels of other characteristics were set to regular. The collected data was stored in the \textit{imitator database} along with the corresponding configuration of GCAT. 

The dominant features were selected from a list of features which consisted of abs\_x, rng\_x, nop\_x, api\_x, bap\_x, rng\_y, nop\_y, eng\_y, sef\_x, bap\_y, sef\_y, nop\_z, bap\_z, sef\_z, mean\_m, api\_m, rng\_m. The abbreviations are as follows: absolute sum (abs), range (rng), number of peaks (nop), energy (eng), spectral edge frequency (sef), band power (bap), and mean (mean). \textit{It is important to note that these features are substantially different from those that have been used for implementing the authentication systems.} Each of these features was examined one by one and added to the dominant feature set (say $\mathbb{D}$) if it was (a) not strongly correlated with any of the features available in $\mathbb{D}$ and (b) strongly correlated with at least one more feature in $\mathbb{L}$ excluding itself. The correlation was considered strong if the | correlation coefficient | $\geq 0.5$. For example, abs\_x would added to $\mathbb{D}$ as it passes both of the conditions as $\mathbb{D} = \phi$. Next, rng\_x would be added to $\mathbb{D}$ because it is not strongly correlated with any features in $\mathbb{D}$ = \{abs\_x\} and is strongly correlated with bap\_x, sef\_y, mean\_m, and api\_m. Similarly, the rest of the features available in $\mathbb{L}$ were examined and added to $\mathbb{D}$ once they passed the stated conditions. The imitator profile helped understand how the changes in a specific GCAT affected the values of the individual dominant features.

Figure \ref{FeedbackAttackProcess} illustrates the imitator training process for mimicking a specific target. The details of the steps are as follows. (1) Steal WSGait from the targeted user. In practice, the targeted user can be enticed to install an App on her phone (see Section \ref{PracticalityAndEffectiveness} for more details). The App will collect the desired sensor readings and send the same to a server accessible to the attackers. (2) Search the imitator database and find WSGait that was the closest (with the least mean absolute error) to the \textit{stolen WSGait}. Take a note of the GCAT configurations corresponding to the \textit{closest WSGait} selected from the imitator database. (3) For each of the five dominant features, conduct the overlap test. A feature passes the overlap test if more than 70\% of its values (extracted from the mimicked samples) fall inside the range of the values extracted from the target's stolen samples. The process terminates if all five dominant features pass the overlap test simultaneously. (4) In case the overlap test fails for a particular feature, the mean error between the values obtained from \textit{closest WSGait} and \textit{stolen WSGait} for that feature is computed and used along with the highest correlation coefficient corresponding to the feature in the imitator profile to generate the next feedback. For example, let the feature $rng\_y$ does not pass the overlap test, the mean error is -3.2, and the correlation between $rng\_y$ and speed was $0.9$, the feedback to the imitator would be to \textit{increase the speed}. (5) With the feedback and GCAT configurations, the imitator produces new patterns. The newly obtained patterns are stored in the imitator database along with the updated GCAT configurations. The steps from (2) to (5) are repeated until every dominant feature simultaneously passes the overlap test.

Three sets of spoof samples were collected after exiting from the feedback loop. Each set consisted of about twenty-two frames (see Section \ref{PreprocessingAndSegmentation}). These frames were then classified using the respective authentication models, and corresponding SFARs were calculated.  

\section{Results and Discussion}
\label{ResultAndDiscussion} 
\subsection{Results} 
Results obtained for Zero- and Treadmill-assisted attack scenarios are described below:
\subsubsection{Zero-effort attack scenario}
\begin{figure}[htp]
\centering
\includegraphics[width=6.2in,height=5.5in]{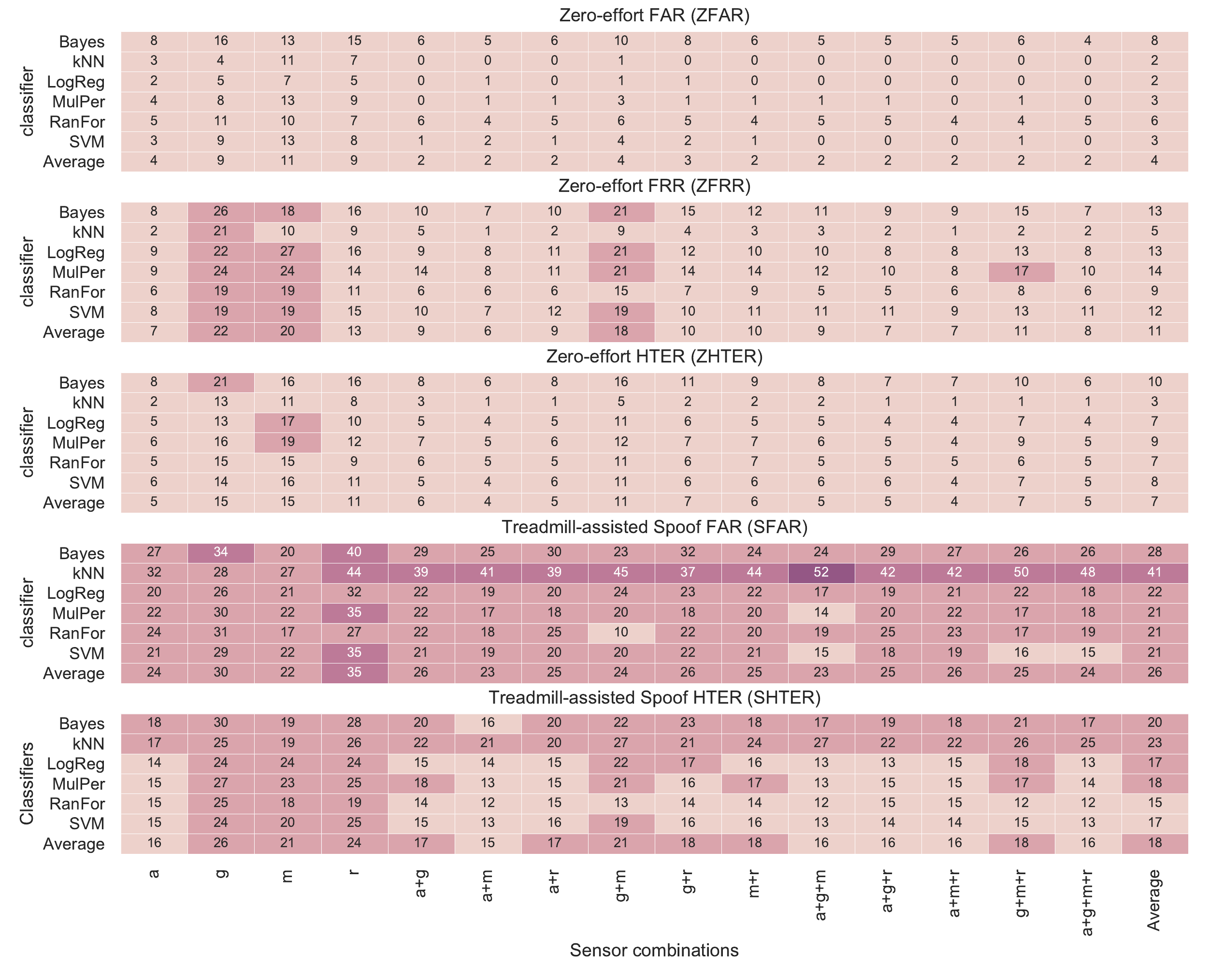}
\caption{a: accelerometer, g: gyroscope, m: magnetometer, and r: rotation vector. The x-axis indicates the combination of sensors. The y-axis represents classification algorithms. Each cell represents percentage error rates for the corresponding combination (classifier and sensors) and is rounded to the nearest integer. A quick comparison between the first (i.e., ZFAR) and the fourth (i.e., SFAR) heatmaps clearly shows the attack's impact. The average FAR has increased from 4\% to 26\% (see the bottom-right corners of ZFAR and SFAR heatmaps). }
\label{AttackImpact}
\end{figure}

The first three heatmaps in Figure \ref{AttackImpact} summarize the performance of WSGait models obtained under the Zero-effort attack scenario. Overall, kNN was the best classifier with the average ZHTER under 3\%, followed by Random Forest and Logistic Regression with ZHTER of 7\%. The rest of the classifiers were able to achieve under 10\% error rates. The obtained error rates either closely matches or better than what has been reported in the past under Zero-effort attack scenarios \cite{PhoneGait7,damavsevivcius2016smartphone,SVMImpBorrowedFrame,PhoneMovementPrinceton,Lee2015MultisensorAT,kNNForGait,muaaz2013analysis, PhoneGait2,muaaz2012analysisWeakDTW,LatestimitationFail,Gafurov2006, GafurovGender2007, Gafurov2007TIFS}. Among the individual sensors (see the first four columns of ZHTER heatmap), the accelerometer has achieved the best error rates, which aligns with the findings of \cite{Lee2015MultisensorAT, PhoneMovementPrinceton}. Also, we observed that WSGait models implemented using multiple sensors exhibited low error rates, in general (compare the first four and last eleven columns of the ZHTER heatmap). Although the magnetometer exhibited very high error rates alone, its combination with the accelerometer achieved the overall best error rate (ZHTER of 4\%). The results concur with the findings of \cite{Lee2015MultisensorAT}.

\subsubsection{Treadmill-assisted attack scenario}
The {fourth} heatmap presents the SFAR, the False Accept Rate obtained under the treadmill-attack scenario. Overall, the SFAR heatmap suggests that WSGait models based on kNN were the most affected (SFAR of 41\%) ones regardless of the sensor combinations used, followed by Bayes Network (SFAR of 28\%). The rest of the classifiers achieved SFAR between 21-22\%. SFAR of more than 20\% renders the WSGait undesirable for many application scenarios. Therefore, we conclude that treadmill-assisted gait spoofing is a threat to WSGait. Besides, the SFAR heatmap suggests that basing a comparison of different design alternatives of WSGait just on the ZFAR, ZFRR, ZHTER, or any metric computed under the Zero-effort attack could be misleading \cite{ChungXu,kNNForGait}.

\subsubsection{Comparison with the previous studies} We compare error rates reported in the previous study and this study in Table \ref{tab:comparision_2015}. We can see that the design alternative presented in this paper is more accurate and robust in both Zero-effort and Treadmill-assisted attack scenarios. It is because of the use of a wide variety of features selected based on mutual information. The SFAR of 22.85\%, however, is problematic for several application scenarios of WSGait.

\begin{table}[htp]
\small
\centering
\caption{Comparison of the performance of the attack on the design of accelerometer-based WSGait presented in this and the previous study \cite{TreadmillAttack}. In the previous study, the authors used a fixed set of seventeen features for implementing all user models. Besides, they used the same set of features to train the imitator. In this work, we use thirty highly discriminative features selected from a pool of 136 time and frequency domain features. In addition, we included the feature selector in the classification pipeline which selected different set of features for each user model making it, intuitively more difficult to breach. The FAR, FRR, and HTER in the Zero-effort columns suggest that the alternative design of WSGait presented in this work not only achieves a reduced (from 6.81\% to 6.14\%) the error rate, but it also makes the WSGait more resilient (current SFAR is 22.85\% compared to the previous SFAR of 38.30\%) to the treadmill-assisted attack. The SFAR of 22.85\% is still very high and makes WSGait undesirable for majority of its application scenarios (see Section \ref{PracticalityAndEffectiveness}).}
\label{tab:comparision_2015}
\begin{tabular}{l|c|c|c|c|c|c|c|c|c|c|}
\cline{2-11}
\multicolumn{1}{c|}{}                                      & \multicolumn{6}{c|}{\textbf{Zero-effort attack (in \%)}}                                              & \multicolumn{4}{c|}{\textbf{Treadmill-assisted attack (in \%)}}             \\ \hline
\multicolumn{1}{|c|}{\multirow{2}{*}{\textbf{Classifier}}} & \textbf{Prev} & \textbf{Curr} & \textbf{Prev} & \textbf{Curr} & \textbf{Prev} & \textbf{Curr} & \textbf{Prev}  & \textbf{Curr}  & \textbf{Prev}  & \textbf{Curr}  \\ \cline{2-11} 
\multicolumn{1}{|c|}{}                                     & \textbf{FAR}  & \textbf{FAR}  & \textbf{FRR}  & \textbf{FRR}  & \textbf{HTER} & \textbf{HTER} & \textbf{SFAR}  & \textbf{SFAR}  & \textbf{SHTER} & \textbf{SHTER} \\ \hline
\multicolumn{1}{|l|}{\textbf{Bayes}}                       & 4.75          & 8.22          & 14.65         & 7.91          & 9.70          & 8.07          & 30.14          & 27.20          & 22.40          & 17.56          \\ \hline
\multicolumn{1}{|l|}{\textbf{LogReg}}                      & 6.16          & 2.07          & 6.46          & 8.87          & 6.31          & 5.47          & 39.16          & 19.91          & 22.81          & 14.39          \\ \hline
\multicolumn{1}{|l|}{\textbf{MulPer}}                      & 5.90          & 3.81          & 6.20          & 8.66          & 6.05          & 6.23          & 40.23          & 22.16          & 23.22          & 15.41          \\ \hline
\multicolumn{1}{|l|}{\textbf{RanFor}}                      & 5.90          & 4.90          & 3.40          & 5.88          & 4.65          & 5.39          & 43.55          & 24.02          & 23.48          & 14.95          \\ \hline
\multicolumn{1}{|l|}{\textbf{SVM}}                         & 5.60          & 2.89          & 9.12          & 8.19          & 7.36          & 5.54          & 38.43          & 20.96          & 23.78          & 14.57          \\ \hline
\multicolumn{1}{|l|}{\textbf{Average}}                     & \textbf{5.66} & \textbf{4.38} & \textbf{7.97} & \textbf{7.90} & \textbf{6.81} & \textbf{6.14} & \textbf{38.30} & \textbf{22.85} & \textbf{23.13} & \textbf{15.38} \\ \hline
\end{tabular}
\end{table}

\subsection{Discussion}
\subsubsection{{Practicality} and effectiveness of the attack}
\label{PracticalityAndEffectiveness} Before we discuss the {practicality} and effectiveness of the attack, we would like to note that the presented attack is proof of concept and can be improved substantially in the future. For example, one can replace a human-plus-treadmill with an electromechanical device for a better calibration of gait characteristics, in turn, the success of the attack \cite{RoboticAttackOnGaitPossibility}. The {practicality} and effectiveness of the attacks, such as the one presented in this work, depends on the application scenario, the amount of information (e.g., information about the target, implementation details of the authentication system, etc.) that the attackers have access to \cite{AttackerTypes} besides the quality of resources (e.g., the quality of treadmill or a sophisticated humanoid robot) deployed, and how much effort is put in, overall \cite{EffortAsAFactorInBiometricAttack}. We describe these factors in the context of the presented-attack as follows.

\textit{Application scenarios--} WSGait may be deployed to protect smart cars \cite{ForbesAppleCarKey}, lockers, etc. or to grant access to critical and highly restricted areas such as secure government offices, cockpit, or military bases \cite{DISAPrototype} as a secondary authentication mechanism if not primary. The possibility of the deployment of gait authentication systems is high because WSGait is considered one of the most challenging traits to mimic \cite{Gafurov2006, GafurovGender2007, Gafurov2007TIFS, StangAttack2007, Mjaaland2011, LatestimitationFail, PredictabilityAndResilienceOfGait}. Items such as smart cars and lockers can be stolen and unlocked at attackers' locations using the presented attack method.

\textit{Required information to launch the attack--} To launch the attack, the attackers need to have access to the gait samples of the target user. The attackers could find numerous ways to steal the required samples. One possible scenario is described below:

The attackers could design an app that collects inertial sensor (requires zero permission \cite{ZeroPermissionSensors2020, ZeroPermissionSesnor}) readings and sends the same to a server administered by them. An invite to install the App can be sent to a list of targeted victims or a list of random victims offering a fantastic shopping deal. The potential victims would install the App to grab the shopping deal. Once attackers receive the target user's data on the server, they can train one or more imitators. Once the attackers are sure that at least one of the imitators is producing the gait patterns that substantially overlap with that of the target's stolen samples, they could launch the presented attack.

\textit{Information and resources that would make the attack easier--} The more information the attackers have access to, the less effort it may take to launch the presented attack. If the attackers know or can estimate height, weight, body type, gender, and age \cite{PrivacyLeakageFromGait2020} of the target user and have access to the implementation details such as features or classification algorithms, the likelihood of the attack success can be improved \cite{GafurovGender2007}. Moreover, the digital treadmill (Make: LifeSpan and Model: TR1200) used in the presented attack is a simple treadmill with standard belt-size and speed control. We believe that advanced treadmills such as the one from NordicTrack (Model: Commercial X32i), which consists of a longer and broader belt, inclination control, push bar, and sled grips would offer better support to the imitators. With more functionalities, support, and control, additional gait factors such as feet or ankle up-down (exploiting inclination) can be defined, which would only help the imitator. 

\subsubsection{Advantage over other attack methods}
\label{AdvantageOverDatainjection} Ratha et al. \cite{AttackSources} have described eight possible attack points in a generic biometrics-based system. All but the first attack point rely on either intercepting the authentication pipeline or overwriting/overriding the stored data. Every attack point, except the first, would be generally guarded by the system-level permissions. In other words, the first attack point relies on the user input; thus, it violates the security principles of "Do not trust user input". Needless to mention that the presented attack exploits the first attack point. Unless we find a way (e.g., liveness detection in fingerprint or face authentication) to assess the input's validity, it would be difficult to thwart the presented attack. 

Although the presented attack requires access to the target's biometric samples, it need not modify the authentication pipeline. In other words, it needs no special privilege. On the other hand, to exploit the rest of the attack points, the attacker would require either a write-permission to modify protected areas of the memory or an interception to a highly encrypted authentication pipeline (likely) or both in addition to access to the target's biometric samples \cite{TypingAttack1, TypingAttack2, ZiboFrogBoiling2012, SwipingAttack3, OneCycleAttack, VideoGANReviewer3}. Researchers have argued that once the attackers get permission to modify the target's device, the race is already over \cite{SerwaddaRobotic, PhoneSwiping1}.

\subsubsection{A numerical explanation of working of the attack?} We conducted an in-depth analysis at the feature level to find a numerical explanation behind the working of the attack. Specifically, we analyzed the feature-wise overlap between the genuine user and impostors (rest of the users in the database) and then genuine and imitator using Bhattacharyya Coefficient ($\mathbb{BC}$), which is defined as follows \cite{BhattaCoeffAndDist}, and have been used for the same purpose in the past \cite{BhattaKeyStroke, Amith2020},

\begin{equation}
\label{BhattacharyyaCoefficient}
\mathbb{BC}({p_1},{p_2})=\sum_{{i=1}}^{n}{\sqrt {p_{1}(i) \times p_{2}(i)}}
\end{equation}

Where, $p_1$ and $p_2$ are two discrete probability distributions over the same domain $X$, $n$ is the number of partitions, $p_{1}(i)$ and $p_{2}(i)$ are the number of data points of the samples $p_1$ and $p_2$ in the $i_{th}$ partition. $\mathbb{BC}$ always lies between zero and one i.e. $0 \le \mathbb{BC}({p_1}, {p_2}) \le 1$. Zero denotes no overlap while one denotes complete overlap between the probability density functions (PDFs) of the two samples.

\begin{table}[htp]
\centering
\caption{Demonstrating the relationship between the inter-class feature-level overlap indicated by Bhattacharyya coefficients $\mathbb{BC}({p_{gen}}, {p_{imp}})$ and the SFARs. We can see that the MedianBCs of 0.63, 0.32, and 0.0 correspond to 61\%, 29\%, and 0\% SFARs, respectively. Note that the authentication models of different user have different features ranked in top thirty (see Section \ref{FeatureAanlysis} for more details).}
\label{tab:BC_High_effort}
\begin{tabular}{|l|l|l|l|l|l|l|}
\hline
\multicolumn{1}{|c|}{{ }}                              & \multicolumn{2}{c|}{{ \textbf{User11 (SFAR = 61\%)}}}                                                         & \multicolumn{2}{c|}{{ \textbf{User18 (SFAR = 29\%)}}}                                                         & \multicolumn{2}{c|}{{ \textbf{User8 (SFAR = 0\%)}}}                                                           \\ \cline{2-7} 
\multicolumn{1}{|c|}{\multirow{-2}{*}{{ \textbf{SN}}}} & \multicolumn{1}{c|}{{ \textbf{Feature}}} & \multicolumn{1}{c|}{{ \textbf{$\mathbb{BC}$}}} & \multicolumn{1}{c|}{{ \textbf{Feature}}} & \multicolumn{1}{c|}{{ \textbf{$\mathbb{BC}$}}} & \multicolumn{1}{c|}{{ \textbf{Feature}}} & \multicolumn{1}{c|}{{ \textbf{$\mathbb{BC}$}}} \\ \hline
{ \textbf{1}}                                          & { npeaks\_x}                             & { 0.25}                                        & { fftc\_std\_dev\_y}                     & { 0}                                           & { stddev\_x}                             & { 0}                                           \\ \hline
{ \textbf{2}}                                          & { mad\_y}                                & { 0.36}                                        & { mad\_y}                                & { 0}                                           & { meanabschange\_x}                      & { 0}                                           \\ \hline
{ \textbf{3}}                                          & { stddev\_m}                             & { 0.39}                                        & { stddev\_y}                             & { 0}                                           & { mad\_x}                                & { 0}                                           \\ \hline
{ \textbf{4}}                                          & { skewness\_z}                           & { 0.45}                                        & { mean\_energy\_y}                       & { 0}                                           & { mean\_energy\_x}                       & { 0}                                           \\ \hline
{ \textbf{5}}                                          & { fquantile\_y}                          & { 0.46}                                        & { fquantile\_z}                          & { 0.04}                                        & { fftc\_std\_dev\_x}                     & { 0}                                           \\ \hline
{ \textbf{6}}                                          & { mean\_energy\_m}                       & { 0.46}                                        & { bin\_counts7\_z}                       & { 0.04}                                        & { fquantile\_x}                          & { 0}                                           \\ \hline
{ \textbf{7}}                                          & { stddev\_y}                             & { 0.5}                                         & { strikeabovemean\_x}                    & { 0.06}                                        & { meanabschange\_y}                      & { 0}                                           \\ \hline
{ \textbf{8}}                                          & { fftc\_std\_dev\_y}                     & { 0.54}                                        & { tquantile\_m}                          & { 0.11}                                        & { strikeabovemean\_z}                    & { 0}                                           \\ \hline
{ \textbf{9}}                                          & { mean\_energy\_y}                       & { 0.55}                                        & { meanabschange\_y}                      & { 0.14}                                        & { ncmean\_y}                             & { 0}                                           \\ \hline
{ \textbf{10}}                                         & { bin\_counts3\_z}                       & { 0.57}                                        & { fftc\_tquantile\_y}                    & { 0.23}                                        & { mad\_y}                                & { 0}                                           \\ \hline
{ \textbf{11}}                                         & { meanabschange\_y}                      & { 0.58}                                        & { fquantile\_y}                          & { 0.23}                                        & { strikeabovemean\_x}                    & { 0}                                           \\ \hline
{ \textbf{12}}                                         & { tquantile\_y}                          & { 0.59}                                        & { mad\_m}                                & { 0.26}                                        & { mean\_energy\_m}                       & { 0}                                           \\ \hline
{ \textbf{13}}                                         & { bin\_counts13\_z}                      & { 0.59}                                        & { meanabschange\_z}                      & { 0.28}                                        & { fftc\_squantile\_y}                    & { 0}                                           \\ \hline
{ \textbf{14}}                                         & { fftc\_std\_dev\_m}                     & { 0.6}                                         & { amean\_m}                              & { 0.3}                                         & { bin\_counts5\_z}                       & { 0}                                           \\ \hline
{ \textbf{15}}                                         & { amean\_m}                              & { 0.63}                                        & { stddev\_m}                             & { 0.3}                                         & { stddev\_y}                             & { 0}                                           \\ \hline
{ \textbf{16}}                                         & { strikebelowmean\_y}                    & { 0.63}                                        & { mean\_energy\_m}                       & { 0.34}                                        & { mean\_energy\_y}                       & { 0}                                           \\ \hline
{ \textbf{17}}                                         & { fquantile\_m}                          & { 0.64}                                        & { fftc\_std\_dev\_m}                     & { 0.38}                                        & { amean\_m}                              & { 0}                                           \\ \hline
{ \textbf{18}}                                         & { npeaks\_z}                             & { 0.72}                                        & { fftc\_squantile\_y}                    & { 0.39}                                        & { tquantile\_m}                          & { 0}                                           \\ \hline
{ \textbf{19}}                                         & { bin\_counts13\_y}                      & { 0.75}                                        & { npeaks\_x}                             & { 0.4}                                         & { ncmean\_z}                             & { 0}                                           \\ \hline
{ \textbf{20}}                                         & { bin\_counts12\_z}                      & { 0.76}                                        & { fquantile\_m}                          & { 0.48}                                        & { tquantile\_y}                          & { 0}                                           \\ \hline
{ \textbf{21}}                                         & { bin\_counts12\_y}                      & { 0.8}                                         & { skewness\_y}                           & { 0.51}                                        & { fftc\_tquantile\_x}                    & { 0.04}                                        \\ \hline
{ \textbf{22}}                                         & { bin\_counts11\_z}                      & { 0.82}                                        & { bin\_counts12\_y}                      & { 0.51}                                        & { tquantile\_x}                          & { 0.06}                                        \\ \hline
{ \textbf{23}}                                         & { npeaks\_y}                             & { 0.85}                                        & { ncmean\_y}                             & { 0.52}                                        & { strikeabovemean\_y}                    & { 0.2}                                         \\ \hline
{ \textbf{24}}                                         & { bin\_counts11\_y}                      & { 0.86}                                        & { tquantile\_y}                          & { 0.58}                                        & { npeaks\_x}                             & { 0.21}                                        \\ \hline
{ \textbf{25}}                                         & { bin\_counts15\_m}                      & { 0.86}                                        & { bin\_counts11\_y}                      & { 0.62}                                        & { strikebelowmean\_x}                    & { 0.25}                                        \\ \hline
{ \textbf{26}}                                         & { npeaks\_m}                             & { 0.89}                                        & { npeaks\_m}                             & { 0.66}                                        & { strikebelowmean\_y}                    & { 0.34}                                        \\ \hline
{ \textbf{27}}                                         & { bin\_counts14\_y}                      & { 0.93}                                        & { bin\_counts13\_y}                      & { 0.67}                                        & { bin\_counts13\_y}                      & { 0.43}                                        \\ \hline
{ \textbf{28}}                                         & { bin\_counts15\_y}                      & { 0.94}                                        & { npeaks\_z}                             & { 0.83}                                        & { fftc\_squantile\_x}                    & { 0.53}                                        \\ \hline
{ \textbf{29}}                                         & { bin\_counts15\_x}                      & { 0.95}                                        & { strikebelowmean\_z}                    & { 0.88}                                        & { skewness\_y}                           & { 0.56}                                        \\ \hline
{ \textbf{30}}                                         & { bin\_counts15\_z}                      & { 0.96}                                        & { npeaks\_y}                             & { 0.93}                                        & { bin\_counts11\_y}                      & { 0.78}                                        \\ \hline
{ \textbf{31}}                                         & { MedianBC}                     & { 0.63}                               & { {MedianBC}}                     & { {0.32}}                               & { {MedianBC}}                     & { {0}}                                  \\ \hline
{ \textbf{32}}                                         & { {MeanBC}}                       & { {0.66}}                               & { {MeanBC}}                       & { {0.36}}                               & { {MeanBC}}                       & { {0.11}}                               \\ \hline
{ \textbf{33}}                                         & { {StdDevBC}}                     & { {0.19}}                               & { {StdDevBC}}                     & { {0.27}}                               & {{StdDevBC}}                     & { {0.21}}                               \\ \hline
\end{tabular}
\end{table}

The higher the feature overlap between the \textit{target} and \textit{impostors/imitator} is the more chances of false accept. Therefore, we computed the $\mathbb{BC}$ for each of the top thirty features extracted from the data collected from \textit{target} and \textit{impostors}, and \textit{target} and the \textit{imitator}. The $\mathbb{BC}$ between the distributions of features of \textit{target} and \textit{impostors} was found to be under $\le 0.30$ for the majority of the top thirty features. While $\mathbb{BC}$ was observed to be mostly $\ge 0.30$ for the substantially affected \textit{targets} and the \textit{imitator}. For reference, we provide the feature overlap for one of the most, average, and least affected \textit{targets'} and the corresponding mimicked samples (see Table \ref{tab:BC_High_effort}). The table suggests that the success of the attack can be roughly estimated by looking at $\mathbb{BC}$ for most features. The median of $\mathbb{BC}$s seems to be a good indicator of the SFAR.

\subsubsection{Is one imitator enough for the high effort attack scenario?} A question arises as to whether a single imitator can mimic any individual's gait patterns. The more in-depth analysis of the user-level errors and demographics of the thirteen genuine users suggested that the users that were in the proximity of the imitator's physical characteristics were affected more, in general. On the other hand, the least affected users were drastically different, in terms of physical traits, from the imitator. The previous study \cite{TreadmillAttack} has reported that while training the imitator for particular users, the training process was terminated for some users in less than ten iterations as the feature overlap passed the criteria. While for some users, the feature overlap did not meet the threshold even after forty iterations. The significant difference between such user and imitator pair was the height and weight. Thus, we posit that the imitators with similar physical characteristics, especially height and weight, would likely be more successful.

\subsubsection{Limitations}
\label{Limitations}Although exhaustive, the experiments presented in this work focused on a widely studied authentication system design that uses frame-based feature extraction and machine learning algorithms. Like most previous studies \cite{LatestimitationFail, Mjaaland2011, StangAttack2007}, this study is also limited in terms of the dataset's size. Although the dataset used in this study was collected in a more realistic than several previous studies, the phone positioning and location were still restricted. In a more realistic scenario, the participant would be allowed to place the phone in the position and location of their wish \cite{OrientationInvariantTIFS2019}. It would be interesting to explore the effectiveness of the attack in that scenario in the future. Furthermore, we would like to note that the data collection process for such investigations is an extraneous exercise for both the researchers as well as participants. It would be interesting to investigate a less mechanical, e.g., an algorithmic way to study and evaluate the robustness of WSGait \cite{ThreatModel}. Nevertheless, this work demonstrated that WSGait could be easy to circumvent and draws the research community's attention toward the problem with WSGait.

\subsubsection{Possible countermeasures} 
Possible countermeasures can include the fusion of sensor reading collected from different devices (e.g., smartphone, smart-watch, and smart-ring). Researchers may also explore methods that would draw a more robust boundary between the genuine and impostor samples.

\section{Conclusion} 
\label{Conclusion} 
\quad The alarming success (average FAR from 4\% to 26\%) of the proof-of-concept attack on authentication systems built upon a variety of wearable sensors calls for further research and development of robust countermeasures and rigorous performance testing before WSGait gets deployed for public use. In the future, we plan to explore a dictionary-based attack and possible countermeasures.
\begin{acks}
We are deeply grateful to the anonymous reviewers for taking the time to review the manuscript and give insightful feedback and comments. 
\end{acks}
%
\bibliographystyle{ACM-Reference-Format}
\bibliography{acmart}
\end{document}